\documentclass[preprint,3p]{elsarticle}

\usepackage{amsmath}
\allowdisplaybreaks

\def \eps{\epsilon}
\def \pipj{(p_i\cdot p_j)}
\def \piq{(p_i\cdot q)}
\def \pjq{(p_j\cdot q)}
\def \ai{\alpha_i}
\def \aj{\alpha_j}
\def \di{d_i}
\def \dj{d_j}
\def \dij{(\di+\dj)}
\def \vm{v_-}
\def \vp{v_+}
\def \lv{\ln(v)}
\def \lvp#1{\ln^{#1}(v_+)}
\def \lni#1{\ln^{#1}\Big(\frac{\ai}{\vp}\Big)}
\def \lnip#1{\ln^{#1}\Big(1-\frac{\ai}{\vp}\Big)}
\def \lnj#1{\ln^{#1}\Big(\frac{\aj}{\vp}\Big)}
\def \lnjp#1{\ln^{#1}\Big(1-\frac{\aj}{\vp}\Big)}
\def \Li{{\rm Li}}
\def \lii#1{{\Li_#1\Big(\frac{\ai}{\vp}\Big)}}
\def \lij#1{{\Li_#1\Big(\frac{\aj}{\vp}\Big)}}
\def \lix#1{{\Li_#1(x)}}
\def \lixp#1{{\Li_#1(1-x)}}

\begin{document}

\title{A simplified expression for the one-loop soft-gluon current with massive fermions}

\author[Aachen]{Micha\l{}  Czakon}
\author[Cambridge]{Alexander Mitov}

\address[Aachen]{Institut f\"ur Theoretische Teilchenphysik und Kosmologie,
RWTH Aachen University, D-52056 Aachen, Germany}
\address[Cambridge]{Cavendish Laboratory, University of Cambridge, CB3 0HE Cambridge, UK}

\date{\today}

\cortext[thanks]{Preprint numbers: Cavendish-HEP-18/08, TTK-18-13}

\begin{abstract}
We present a much simpler analytic expression for the UV un-renormalized one-loop soft-gluon current which controls the infra-red behavior of one-loop QCD amplitudes with massive fermions when an external gluon becomes soft. The result is given entirely in terms of standard polylogarithms, is of uniform transcendentality and is well-suited for computer implementation.
\end{abstract}
\maketitle

In ref.~\cite{Bierenbaum:2011gg} the expression for the one-loop correction to the so-called soft-gluon current \cite{Catani:2000pi} with massive fermions was derived. This process-independent result controls the singular behavior of one-loop QCD amplitudes with massive fermions when one of the external gluons becomes soft. It is also an essential ingredient to NNLO QCD calculations for processes with massive fermions.

For one-loop amplitudes with a single massive fermion leg the analytic expression for that current is quite compact. It was re-derived in ref.~\cite{Brucherseifer:2013iv}. For one-loop amplitudes with two massive fermion legs the soft-gluon current through order ${\cal O}(\eps^0)$, where $\eps=(4-d)/2$ is the dimensional regularization parameter, is also simple and is given in terms of standard polylogarithms. The term of order ${\cal O}(\eps)$, however, is quite large in size and contains a function which is not written through polylogarithms. 

In this work we rewrite the result of ref.~\cite{Bierenbaum:2011gg} in an equivalent but much more compact form which contains only standard polylogarithms of up to weight three. The result is given in a form that is particularly well suited for computer implementation. 

Following ref.~\cite{Bierenbaum:2011gg}, the one-loop correction to the UV un-renormalized soft-current is expressed through the function
\begin{equation}
g^{(1)}_{ij} = - \frac{1}{2} \, a^b_S \, \Bigg( \frac{2 \pipj \mu^2}{2
  \piq 2 \pjq} \Bigg)^\eps \, \Bigg[ \frac{1}{\eps^2} + \sum_{n=-1}^1
\eps^n \Big( R_{ij}^{(n)} + i\pi I_{ij}^{(n)} \Big) \Bigg] \; ,
\label{eq:gij}
\end{equation}
where the fixed indices $i,j$ label the two massive legs with momenta $p_i, p_j$ such that $p_i^2=m_i^2$ and $p_j^2=m_j^2$. We note that the function $g^{(1)}_{ij}$ in eq.~(\ref{eq:gij}) corresponds precisely to the function $g^{(1)}_{ij}$ in ref.~\cite{Bierenbaum:2011gg}.

We give the result for the function $g^{(1)}_{ij}$ for both time-like ($TL$) and space-like ($SL$) kinematics. In the time-like case both momenta $p_i$ and $p_j$ are outgoing. This configuration corresponds to {\it Case~3} of ref.~\cite{Bierenbaum:2011gg}. The case of both $p_i$ and $p_j$ incoming is obtained by changing the sign of the imaginary part. The space-like case corresponds to $p_i$ outgoing and $p_j$ incoming. Its real part can be expressed through the real part of the time-like case as follows
\begin{equation}
\begin{aligned}
R_{ij}^{(-1) [SL]} =& \, R_{ij}^{(-1) [TL]} \; , \\[0.4cm]
R_{ij}^{(0) [SL]} =& \, R_{ij}^{(0) [TL]} -12 \zeta_2 \frac{\vm}{v} \;
, \\[0.2cm]
R_{ij}^{(1) [SL]} =& \, R_{ij}^{(1) [TL]} + 12 \zeta_2 \frac{1}{v}
\Bigg[ \frac{2}{\dij} \big( \aj \vp - \ai \vm \big) \lnj{} + \big(
\vp \lvp{} - \lv \big) \Bigg] \; ,
\end{aligned}
\end{equation}
while its imaginary part can be written explicitly as
\begin{equation}
\begin{aligned}
I_{ij}^{(-1) [SL]} =& \, 1 \; , \\[0.2cm]
I_{ij}^{(0) [SL]} =& \, \frac{2}{v \dij} \Big( \big( \ai - \vm \big)
\lni{} - \big( \di \vm + \aj v \big) \lnj{} \Big) + \lvp{} \; , \\[0.2cm]
I_{ij}^{(1) [SL]} =& \, \frac{1}{\dij} \Bigg\{ \big( 1 - \dij \big)
\Bigg[ 2 \lni{} \lnip{} - \lnj{} \Big( 2 \lnjp{} + \lvp{} \Big) \\&
- 2 \lij2 + 2 \lii2 \Bigg] + \frac{1}{v} \Bigg[ \big( \ai - \vm \big)
\lni2 + \big( \di \vm + \aj v \big) \lnj2 \\&
+ 2 \lni{} \Big( \big( \vp - \ai \big) \lvp{} - \di \lv \Big) + \di
\lnj{} \Big( \lvp{} - 2 \lv \Big) - 2 \di \lix2 \\&
- 2 \zeta_2 \dj \Bigg] \Bigg\} + \frac{1}{2} \lvp2 - \zeta_2 \Big(
\frac{3}{2} - \frac{2}{v} \Big) \; .
\end{aligned}
\end{equation}

The corresponding time-like results read:

\begin{equation}
\begin{aligned}
R_{ij}^{(-1) [TL]} =& \, \lvp{} - \frac{\vm}{v} \Big(\lni{} + \lnj{}\Big) \; ,
\\[0.2cm]
R_{ij}^{(0) [TL]} =& \, \frac{1}{v} \Bigg[ \frac{1}{\dij} \Big( (\ai \vp - \aj
\vm) \lni2 + \big( \aj \vp - \ai \vm \big) \lnj2 \Big) \\&
+ \Big( \lni{} + \lnj{} \Big) \big( \vp \lvp{} - \lv \big) - \lix2
\Bigg] + \frac{1}{2} \lvp2 + \zeta_2 \Big( \frac{7}{v} - \frac{19}{2}
\Big) \; , \\[0.2cm]
R_{ij}^{(1) [TL]} =& \, \frac{1}{\dij} \Bigg\{ \big (1 - \dij \big) \Bigg[
\lnip{} \lni2 + \lnjp{} \lnj2 \\&
+  2 \Big( \lni{} \lii2 + \lnj{} \lij2 \Big) - \lix2 \Big( \lni{} +
\lnj{} \Big) \\&
+  2 \Big( \lix3 - \lii3 - \lij3 + \zeta_3 \Big) \Bigg] - 7 \zeta_2
\Big( \lni{} + \lnj{} \Big) \\&
+ \frac{1}{v} \Bigg[ \Big( \big(\aj \vp - \ai \vm \big) \lni2 + \big(
\ai \vp - \aj \vm \big) \lnj2 \Big) \lvp{} \\&+ \big( \ai - \aj \big)
\Big( \lni2 - \lnj2 \Big) \lv - \Big( \di \lni{} + \dj \lnj{} \Big)
\big( \lix2 - 7 \zeta_2 \big) \Bigg] \Bigg\} \\&
+ \frac{1}{v} \Bigg\{ \Bigg[ \lvp{} \Big( \frac{3 + v}{4} \lvp{} - \lv
\Big) - \frac{9 \vm}{2} \zeta_2 \Bigg] \Big( \lni{} + \lnj{} \Big) \\&
-\frac{\vm}{6} \Big( \lni3 + \lnj3 \Big) + 2 \lixp3 + \lix3 - \Bigg[ \lix2 +
\zeta_2 \Big( 5 + \frac{19}{2} v \Big) \Bigg] \lvp{} \\&
+ 12 \zeta_2 \lv \Bigg\} + \frac{1}{6} \lvp3 - \Big( \frac{7}{3} +
\frac{1}{v} \Big) \zeta_3 \; ,
\end{aligned}
\end{equation}
and
\begin{equation}
\begin{aligned}
I_{ij}^{(-1) [TL]} =& \, 2 - \frac{1}{v} \; , \\[0.2cm]
I_{ij}^{(0) [TL]} = & \, \frac{2}{v} \Bigg[ \frac{1}{\dij} \Big( \big(
\ai - \vm \big) \lni{} + \big( \aj - \vm \big) \lnj{} \Big) + \Big(
\frac{1}{2} + v \Big) \lvp{} - \lv \Bigg] \; , \\[0.2cm]
I_{ij}^{(1) [TL]} =& \, \frac{1}{\dij} \Bigg\{ 2 \big( 1 - \dij \big)
\Big( \lni{} \lnip{} + \lnj{} \lnjp{} 
+ \lii2 \\&
+ \lij2 \Big) + \Big( \lni{} + \lnj{} \Big) \lvp{} - 2 \big( \lix2 +
\zeta_2 \big) + \frac{1}{v} \Bigg[ \Big( \big( \ai - \vm \big) \lni2
\\&
+ \big( \aj - \vm \big) \lnj2 \Big) + \Big( \di \lni{} + \dj \lnj{}
\Big) \big( \lvp{} - 2 \lv \big) \Bigg] \Bigg\} \\&
- \frac{1}{v} \Big( 4 \vm \lix2 + \frac{1}{2} \big( \lvp{} - 2 \lv
\big)^2 \Big) + \lvp2 - \zeta_2 \Big( 1 - \frac{3}{2 v} \Big) \; .
\end{aligned}
\end{equation}

The above results have been expressed through the following variables
\begin{equation}
\begin{gathered}
\ai \equiv \frac{m_i^2  \, 2 \pjq}{2 \pipj 2 \piq} \; , \quad
\aj \equiv \frac{m_j^2 \, 2 \piq}{2 \pipj 2 \pjq} \; , \quad
\di \equiv 1 - 2 \ai \; , \quad \dj \equiv 1 - 2 \aj \; , \\[0.4cm]
v \equiv \sqrt{1 - 4 \ai \aj} \; , \quad v_{\pm} \equiv \frac{1 \pm
  v}{2} \; , \quad  x \equiv \frac{\vm}{\vp} \; ,
\end{gathered}
\end{equation}
and $\zeta_n$ is the Riemann zeta function. For ease of comparison with ref.~\cite{Bierenbaum:2011gg} we have retained as much as possible the notation introduced there. We only note that the variables $\ai$ and $\aj$ used here differ from their definition in ref.~\cite{Bierenbaum:2011gg} by a factor of $1/2$, the normalization of the functions $R_{ij}$ and $I_{ij}$ differs from the one in ref.~\cite{Bierenbaum:2011gg} by the overall factor of $-1/2$ in eq.~(\ref{eq:gij}) and the variable $x$ used here corresponds to $x^2$ from ref.~\cite{Bierenbaum:2011gg}.

The above result for the UV un-renormalized soft-gluon current is of uniform transcendentality. It also has the nice property that the limits $m_i \to 0$ and/or $m_j \to 0$ are smooth. If $m_i = 0$, the correct result is obtained by setting $\ln(\ai/\vp) = 0$ followed by $\ai = 0$. Similarly for $m_j = 0$. In either case $v = v_+= 1$ and $v_- = x = 0$. 

The UV renormalization of the results presented here is trivial; see ref.~\cite{Bierenbaum:2011gg} for details.

\noindent
\section*{Acknowledgments}
A.M. thanks the Department of Physics at Princeton University for hospitality during the completion of this work. The work of A.M. is supported by the UK STFC grants ST/L002760/1 and ST/K004883/1 and by the European Research Council Consolidator Grant NNLOforLHC2.

\end{document}